%% file: paper.tex
\documentclass[sigconf]{acmart}

\usepackage{mdframed}
\usepackage{makecell}
\usepackage{enumerate}

\newtheorem{principle}{Principle}

\title{Single versus Double Blind Reviewing at WSDM 2017}

\author{Andrew Tomkins}
\affiliation{%
\institution{Google}
\city{Mountain View}
\state{CA}
\country{USA}
}
\email{atomkins@gmail.com}
\author{Min Zhang}
\affiliation{%
\institution{Tsinghua University}
\city{Beijing}
\country{China}
}
\email{z-m@tsinghua.edu.cn}
\author{William D.\ Heavlin}
\affiliation{%
\institution{Google}
\city{Mountain View}
\state{CA}
\country{USA}
}
\email{bheavlin@google.com}

\newif\ifformatforarxiv
\formatforarxivtrue
\ifformatforarxiv
  \renewcommand\footnotetextcopyrightpermission[1]{} 
\fi

\begin{document}

\acmConference[KDD'17]{Knowledge Discovery and Data Mining}{August 13--17, 2017}{Halifax, Nova Scotia, Canada}

\begin{abstract}
\input abstract
\end{abstract}

\maketitle

\input{intro}

\input{rel}
\input{experiment}

\input{data}
\input{analysis}
\input{meta}
\input{conclusion}


\bibliographystyle{abbrv}
\bibliography{refs}

\input{appendix}

\end{document}

%% file: abstract.tex
  In this paper we study the implications for conference program
  committees of using single-blind reviewing, in which committee
  members are aware of the names and affiliations of paper authors,
  versus double-blind reviewing, in which this information is not
  visible to committee members.  WSDM 2017, the 10th ACM International
  ACM Conference on Web Search and Data Mining, performed a controlled
  experiment in which each paper was reviewed by four committee
  members.  Two of these four reviewers were chosen from a pool of
  committee members who had access to author information; the other
  two were chosen from a disjoint pool who did not have access to this
  information.  This information asymmetry persisted through the
  process of bidding for papers, reviewing papers, and entering
  scores.  Reviewers in the single-blind condition typically bid for
  22\% fewer papers, and preferentially bid for papers from top
  institutions.  Once papers were allocated to reviewers, single-blind
  reviewers were significantly more likely than their double-blind
  counterparts to recommend for acceptance papers from famous authors
  and top institutions.  The estimated odds multipliers are 1.63 for
  famous authors and 1.58 and 2.10 for top universities and companies
  respectively, so the result is tangible.  For female authors, the
  associated odds multiplier of 0.78 is not statistically significant
  in our study.  However, a meta-analysis places this value in line
  with that of other experiments, and in the context of this larger
  aggregate the gender effect is also statistically significant.

%% file: intro.tex
\section{Introduction} \label{sec:intro}

The scientific peer-review process dates back to the 1600's, and is
generally regarded as a cornerstone of the scientific method.  The
details of its implementation have been scrutinized and explored
across many academic disciplines.  

Peer review has many dimensions.  At the present time, there is a
conversation underway throughout the scientific community regarding
\emph{open peer review}, which covers a range of practices ranging
from revealing reviewer names to authors to making peer reviews
available to the general public, with or without the reviewer's name
attached.  In the last 2--3 decades there have been numerous trials of
open peer review.  Perhaps most visible is a long-running experiment
by the journal \emph{Nature}.  After lukewarm initial
experiences~\cite{nature06}, a decade later \emph{Nature} now reports
60\% of reviewers are comfortable with their reviews becoming public,
given the right of the reviewer to withhold his or her
name~\cite{nature16}.  While this important trend continues to
generate lively discussion, we do not discuss open peer review further
in this paper.

Rather, our focus is on the question of availability to reviewers of
information about the authors.  This question remains an active area
of debate, with many significant conferences and journals on each side
of the question.  Terminology is not completely uniform across the
sciences, but following common usage in computer science, we refer to
\emph{single-blind reviewing} as the practice of making reviewers
aware of author identity but not the other way around.  In
\emph{double-blind reviewing}, neither party is aware of the identity
of the other.

Numerous anecdotal studies argue for one form or the other of peer
review, often based on observations of findings before and after
switching models.  A much smaller number of researchers have performed
controlled studies of the effects of the two models.  Notable among
these is the work of Rebecca Blank from 1991~\cite{blank91}, who
performed a beautiful controlled study in reviewing papers submitted
to the American Economic Review over a two-year period from 1987 to
1989.  We discuss this and other related work in some detail
below.

The current work came about when two of the authors of this paper were
asked to co-chair the program of WSDM 2017, the 10th International ACM
Conference on Web Search and Data Mining.  WSDM has for its entire
history employed single-blind reviewing.  We were asked to consider
switching to double-blind this year.  Upon a review of the literature,
we discovered that earlier controlled experiments in the journal
setting missed many key aspects of the standard WSDM reviewing
process, while many discussions of conferences switching between
reviewing methods were uncontrolled experiments in the sense that the
switch took place from one year to the next, introducing an
analytically intractable set of possible confounding factors.  Hence,
we decided to perform an experiment in order to make an informed
recommendation to the chairs of WSDM 2018, and to offer our findings
to the rest of the community.

We now summarize some differences between conference and journal
reviewing processes.  As a backdrop, we observe that the accelerated
pace of computer science in recent decades has led to the ascendance
of academic conferences as a primary means for dissemination of new
results.  The level of formal methodological scrutiny applied to the
conference paper acceptance process is therefore lower than it is for
peer-reviewed journals.  Some elements that are common in the process
of conference reviewing are less common in a journal review setting,
for instance:
\begin{itemize}
\item Conference review processes often run on an annual cycle, which
  results in large number of papers being reviewed by a large pool of
  reviewers on a single operating schedule.
\item As a result, many conferences operate at a scale that makes it
  difficult for each paper to be matched by an expert to expert
  reviewers.
\item The assignment of reviewers to papers is therefore performed
  using other mechanisms.  In many cases, reviewers are asked to
  indicate ability or interest in reviewing each paper as input to the
  assignment process.  This process is referred to as \emph{bidding}.
\item Each reviewer typically reviews a batch of papers, with a single
  deadline for completing all reviews.
\item Final decisions are often made with constraints on the overall
  number of slots, rather than on a notional quality standard.
\item Decisions are accept or reject; there is typically no option for
  re-review after revision.\footnote{Some conferences add a
    ``rebuttal'' or ``author feedback'' round after reviewing is
    complete to allow authors to respond to reviewer comments or
    describe changes that will appear in the final revision.
    Additionally, if a strong paper has an addressable flaw, some
    conferences will ``accept with shepherding'' appointing a
    knowledgeable party to verify that the flaw has been addressed.
    WSDM does not use rebuttal, and very rarely uses shepherding.}
\end{itemize}

These differences are not hard and fast rules, but the conference
setting does raise different questions about best practices.

In the WSDM setting, we find significant differences between
single-blind and double-blind reviewing.  First, we find that
single-blind reviewers enter about 22\% fewer bids on average, a
highly significant decrease (Mann-Whitney U, $p=.0002$).  We show
that, given these fewer bids, there is a significant preference to bid
on papers from top universities and companies, compared to
double-blind reviewers ($p=0.011$ and $p=0.010$ respectively).

Once the bids have been received, papers are allocated to reviewers,
and we may study the resulting review scores.  We find that the
likelihood for single-blind reviewers to enter a positive review is
significantly higher for papers with a famous author ($p=0.027$) and
for papers from a top university ($p=0.012$) or a top company
($p=0.002$) compared to double-blind reviewers.  The estimated odds
multipliers are 1.63, 1.58 and 2.10 respectively, equivalent to
increases in underlying quality of 0.57 to 0.92 standard deviations.
The effect is strong enough to warrant serious discussion on the
appropriate reviewing policy.

Our findings with respect to bidding imply that reviewers bid less
under single-blind reviewing.  This reduced bid landscape may result
in a lower--quality allocation of papers to knowledgeable reviewers.
It may also be disadvantageous if it results in an ``unfair'' bidding
pool, in favor of papers from top institutions.  It is an ethics and
policy question to determine whether a reviewer who (let us imagine)
implicitly uses information about the quality of the paper's
institution to estimate that the paper is more interesting, and hence
enters a more positive bid on that paper, is acting in a manner that
should be discouraged.

Our findings with respect to reviewing raise similar questions.  A
reviewer who knows that a particular paper is from a top school, or
has a famous author, is significantly more likely to recommend
acceptance than a reviewer who does not know this information.  There
are at least two points for consideration here.  The first is that the
two reviewers are not identical: reviewers that bid on a paper are
more likely to be assigned to review the paper, and as we have already
discussed, the bidding dynamics of the two reviewers are different.
Hence, it is possible that the single-blind reviewer of a particular
paper may have bid on the paper due to knowledge of the author's prior
work, while the double-blind reviewer may have bid due to the topic of
the paper implied by the title.  In other words, paper assignment in
not random, and this should be taken into account in interpreting our
findings.

Our second point with respect to reviewing is that, whatever the
process that resulted in the reviewers being assigned the paper, the
single-blind reviewers with knowledge of the authors and affiliations
are much more positive regarding papers from famous authors and top
institutions.  Again the implications are not cut and dried, but it is
reasonable to raise the concern that authors who are not famous and
not from a top institution may see lower likelihood for acceptance of
\emph{exactly the same work.}

In Section~\ref{sec:meta}, we perform a meta-analysis of our study
along with six other experimental studies from the literature.  Our
findings in this meta-analysis are as follows.  First, with respect to
famous authors, our effect is in fact smaller than the aggregate.  For
top 50 institutions, only one study covered this effect, and showed a
smaller effect than ours.  With respect to female authors, our effect
ranks 7th out of 11 measurements, but is not qualitatively different
than that observed by other authors.  By the standards of
meta-analysis, in aggregate, the effect against female authors can be
considered statistically significant.

We therefore recommend for upcoming WSDM conferences that the program
chairs strongly consider moving to an overall policy of double-blind
reviewing.

%% file: rel.tex
\section{Related Work} \label{sec:rel}

There is extensive literature on scientific peer reviewing overall,
and on single-blind versus double-blind reviewing in particular.  A
survey of Snodgrass~\cite{snodgrass2006} reviews over 600 separate
pieces of literature on reviewing.

For detailed information, we refer to reader to the excellent survey
of Snodgrass~\cite{snodgrass2006} and additional survey material
referenced there.  For editorial perspective, see
Snodgrass~\cite{snodgrass2007editorial},
McKinley~\cite{mckinley2008improving,mckinley15}, or for an argument that the
benefits of double-blind reviewing are small compared to the costs,
see Schulzrinne~\cite{schulzrinne2009double}.

Regarding the peer reviewing landscape, Walker and Rocha da
Silva~\cite{walker15} argue that the soft sciences more commonly apply
double-blind review, while in the natural sciences single-blind review
is more common.  Multiple journals and conferences are moving to
double-blind review (see below for some results studying the
change\-over in those venues).  In some cases the movement is in the
other direction, for example the American Economic Association
announced in 2011 that it will end double-blind review, citing
difficulty of maintaining anonymity and decreased information for
reviewers (for example, to assess authors' possible conflicts of
interest)~\cite{jaschik11}.

\subsection{Specific biases described in the literature}

A number of specific biases are described in the literature.  We will
focus here on literature related to three biases we consider in our
study: gender, prestige, and institution of the authors.

Knobloch-Westerwick \emph{et al.}~\cite{matilda13} propose the
\emph{Matilda effect}, in which papers from male first authors are
evaluated to have greater scientific merit than papers from female
first authors, particularly in male-dominated fields.  The authors
study this effect by randomly assigning names to conference abstracts,
and then asking study participants for assessments of merit.  Studies
disagree on the presence of gender bias in reviewing: many studies
find effects, but many of these are retrospective studies of venues
that have moved to double-blind reviewing, and hence cannot rule out
the possibility that the findings are due to improving gender equality
rather than the reviewing model itself.  Our study does not find a
statistically significant gender effect.  However, in our field there
are no fixed conventions for ``first authors'' so we simply study the
presence of a female author on the paper, which may weaken the effect.
Blank's study~\cite{blank91} shows a small difference depending on
first author gender, but the data volume is low and the difference is
not statistically significant.

Work by Robert Merton~\cite{matthew68} in 1968 proposed the
\emph{Matthew effect}, in which already-famous researchers receive the
lion's share of recognition for new work.  The paper provides an
enlightening survey of this effect through contemporary and historical
science, and cites various psychological processes that may be at
work.  There has been significant follow-on work in this area; see for
instance the discussion of reviewing at the ACM SIGMOD conference
below.

Finally, Blank's study~\cite{blank91} spends significant time
discussing biases resulting from the fame or quality of the authors'
institution(s).

\subsection{Retrospective studies}

In 2001, the journal \emph{Behavioral Ecology} switched from
single-blind to double-blind review.  Budden \emph{et
  al.}~\cite{budden2008} describe their findings analyzing data before
and after the switch.  They found an increase in female first-authored
papers after the change.  Webb \emph{et al.}~\cite{webb2008}, however,
argue that comparable journals that did not switch reviewing model
also showed such an increase over a similar time period.

Roberts and Verhoef~\cite{evolang16} study double-blind reviewing at
the Evolution of Language conference series, comparing the results in
2016, which used double-blind reviewing, to the results of 2012 and
2014, which used single-blind reviewing.  The authors showed a
significant effect for gender, in which papers with female first
authors and male first authors were accepted with similar likelihood
under single-blind reviewing, but female first-author papers were
accepted with higher likelihood under double-blind reviewing.

In 2001, the ACM SIGMOD conference on management of data moved to
double-blind reviewing.  After five years in the new model, Madden and
DeWitt~\cite{madden06} asked whether double-blind reviewing helped
junior researchers who might have been disadvantaged under
single-blind reviewing.  They studied the acceptance rate of more
senior authors before and after the reviewing change.  Their study
showed no difference on acceptances before and after the reviewing
change. However, a follow-on study by Tung~\cite{tung06} analyzing the
same data using a more standard statistical test showed the opposite
result.

\subsection{Experimental studies}


Peters and Ceci~\cite{peters82} performed a notorious study of
reproducibility of peer review results.  The authors of the study
asked for and received permission from the authors of twelve
prestigious papers to re-submit these papers to the journal in which
they appeared, introducing false author names and referencing
manufactured low-prestige institutions (\emph{e.g., the ``Northern
  Plains Center for Research''}).  3 of 38 editors and reviewers
detected the re-submission, so only 9 of the papers were reviewed
fully.  Of those, 8 were rejected, often citing serious methodological
flaws.  In addition to raising concerns about the ethics of peer
reviewing practices, the study itself gained additional notoriety in
part because an ethical debate arose regarding the propriety of the
methodology; the authors provide an insightful discussion of the
history~\cite{peters14}.

The study of Peters and Ceci was published with significant commentary
from many fields, and is frequently referenced in policy discussions.
In addition to the authors' original intent of understanding the
importance of reputation in acceptance decisions, the findings also
raised questions about the overall reproducibility of acceptance
decisions.  Rothwell and Martyn~\cite{rothwell00} went on to study
this question, and found in their setting that reviewers did not agree
with one another regarding a manuscript better than random chance
would indicate.  In Computer Science, the Neural Information
Processing Systems (NIPS) conference subsequently ran an experiment in
which a subset of papers were sent through two parallel review
processes.  Their findings~\cite{nips1, nips2} show that, if the
committee were to re-select papers again, 38--64\% of the papers would
have been accepted again.  We discuss this question in
Section~\ref{sec:agreement}.

Perhaps the best known experimental study of single-blind versus
double-blind reviewing behavior, and to our knowledge the only
controlled experiment in this area other than our own, is the study of
Rebecca Blank~\cite{blank91}.  Over several years, 1498 papers were
randomly assigned to single-blind versus double-blind reviewing
condition.  While Blank performs detailed analyses of many facets of
the data, we may summarize part of the high-level findings as
follows.  First, authors at top or bottom institutions do not see
significant differences in acceptance decisions based on reviewing
model, but authors at mid-tier institutions perform better in a
single-blind setting, as do foreign authors and those outside
academia.  Second, there is a mild indication, not statistically
significant, that women do slightly better in double-blind review.

Recently, Okike \emph{et al.}~\cite{jama16} performed an ingenious
study constructing an artificial submission proposing a study of the
efficacy of training to improve communication in the operating room.
The fabricated study was submitted to an Orthopaedics journal, and
listed as authors two past presidents of the American Academy of
Orthopaedic Surgeons.  With the involvement of the journal, the study
was sent to 256 reviewers, of whom 119 completed the review, split
between single-blind and double-blind conditions.  The results showed
that single-blind reviewers were significantly more favorable toward
the paper.

\subsection{Difficulties implementing double-blind reviewing}

Hill and Provost~\cite{hill03} study the problem of automatically
identifying the authors of a double-blind paper.  They show fully
automated techniques to identify authors with 40--50\% accuracy, and
80\% accuracy for highly prolific authors with 100 or more prior
publications.  Section~\ref{sec:impl} discusses this issue in more
detail.

%% file: experiment.tex
\section{Experiment} \label{sec:experiment} \label{sec:exp}

In this section we describe the design of our experiment.  We begin
with an overview of the reviewing process WSDM has typically employed
in the past:
\begin{enumerate}
\item Program chairs invite program committee (PC) and senior program
  committee (SPC) members while authors submit papers.
\item PC and SPC members bid on each paper, specifying which are of interest.
\item Program chairs perform an assignment of 3--4 PC members and one
  SPC member to each paper, typically resulting in 6--10 papers
  assigned to each PC member.
\item PC members complete reviews of assigned papers.
\item For each paper, the assigned SPC member conducts a discussion
  with the PC members reviewing the paper and makes a recommendation
  for or against acceptance.
\item Based on all this information, the program chairs make final
  decisions.
\end{enumerate}

\subsection{Ethical Considerations in Designing the Experiment} 
\label{sec:ethics}

We spent significant time in discussion about the most appropriate
design for our experiment, given the many ethical considerations, and
we were fortunate to receive valuable input and discussions from the
conference general chairs, the WSDM steering committee, and the Ethics
Committee for Information Sciences (ECIS) at the University of Amsterdam
and the VU University Amsterdam.  

Through this discussion, we adopted two ethical principles in our
design of the experiment:
\begin{principle}
\textbf{No-Bias Condition} A paper's likelihood of acceptance should not
change based on its experimental condition.
\end{principle}

\begin{principle}
\textbf{Veracity Condition} We will not lie to any participant in the
experiment.
\end{principle}

The first principle in particular put significant constraints on
possible experimental designs, as described in
Section~\ref{sec:design}.

Our Call for Papers~\cite{cfp} asks authors to submit PDF documents
that have been anonymized by removing references to the authors and
their institutions.  The CFP does not commit to a particular reviewing
model.  The relevant section reads as follows: ``As an experiment this
year, WSDM 2017 will use a combination of single-blind reviewing and
double-blind reviewing.  Please contact the PC chairs at the address
below for any questions on the submission or review process.''


\subsection{Design of Experiment} \label{sec:design}

We did not see an experimental design that tested the end-to-end
decision process in a way that is consistent with the two ethical
principles above.  Hence, we ran the experiment through the end of the
PC reviewing phase, and terminated the experiment before beginning the
discussion or final decision phases.  The experiment considered only
the behavior of the PC, not the SPC.  Our findings therefore relate
just to bidding, reviewing, and scoring by PC members.  The
experimental design is described in Figure~\ref{fig:exp-design}.
\begin{figure}
\begin{mdframed}
\begin{enumerate}
\item Program committee is split randomly into two groups of equal
  size:  single-blind PC (SBPC) and double-blind PC (DBPC).
\item During bidding, SBPC see author names and affiliations, while
  DBPC do not.  Both groups see paper titles and abstracts.
  Otherwise, the bidding interface is the same.
\item A separate assignment is computed for SBPC and DBPC using the
  standard assignment algorithm provided by the EasyChair conference
  management system.  The overall assignment allocates 4 PC members to
  each paper with exactly 2 from SBPC and 2 from DBPC.
\item The assigned papers are sent for reviewing.  SBPC and DBPC again
  receive the same reviewing form, except that SBPC members see author
  names and affiliations in the reviewing form.  PDF documents do not
  include author names or affiliations.
\item After reviews are received, the experiment is closed, and the
  data is set aside for analysis.  From this point forward, all PC
  members operate in a single-blind condition.  Discussions are
  managed by the SPC member assigned to each paper, in which all 4 PC
  members are now able to see the author names and affiliations for
  the paper.
\end{enumerate}
\end{mdframed}
\caption{Experiment Design} \label{fig:exp-design}
\end{figure}

Note that our goal is not to determine whether a particular paper is
more likely to be accepted in single-blind or double-blind reviewing.
The variance in any single decision is too large to measure this
directly (see Section~\ref{sec:agreement} for some elaboration on this
point).  Instead, we wish to measure statistical differences in the
overall behavior of SBPC and DBPC.  These differences may be measured
in the context of particular classes of papers (papers from top-tier
institutions, papers with female authors, etc) or particular classes
of paper/reviewer pairs (reviewer from the same country as the paper,
etc).

Due to the design, we may study the impact of single-blind versus
double-blind reviewing on the bidding process, and on how reviewers
score papers.  We can show whether papers written by a particular
gender are more likely to receive bids and more likely to receive
higher review scores in the single-blind or double-blind condition.
We cannot see how the reviewing model impacts SPC recommendations or
final paper acceptances.  However, we felt that if significant
behavioral differences exist, we should observe this in the
experiment.

We considered and rejected a number of alternative approaches to the
experiment, including the following:
\begin{enumerate}
\item Splitting papers between a single-blind and a double-blind condition.
We rejected this approach because authors could reasonably argue that
being placed in a particular condition could have reduced their
likelihood of acceptance.
\item Splitting each reviewer into some single-blind and some
  double-blind reviews.  We rejected this approach because it is not
  well-defined how to perform bidding in this setting, and also
  because it would implicitly force reviewers to compare their
  behavior with respect to the two groups of paper, which might
  introduce biases.

\item Removing reference to the experiment from the CFP and our
  communications with reviewers.  We rejected this approach because we
  felt it would entail at some level lying to both authors and
  reviewers about the process.
\item Sending a small number of papers through both a single-blind and
  double-blind condition in parallel.  We performed rough calculations
  to infer that we would not have sufficient statistical strength in
  this approach to make clean statements about the outcomes.  We also
  were concerned that any reasonable scheme to fuse the results of the
  two decision processes would be inconsistent with our no-bias
  principle.
\end{enumerate}

%% file: data.tex
\section{Data} \label{sec:data}

In this section, we describe the data available from our experiment.
Each reviewer in the experiment performed two tasks: bidding for
papers, and then reviewing a set of assigned papers.

During bidding, each reviewer considered the submitted papers and
entered a bid for each.  Three bids are possible: yes, maybe,
no.\footnote{There is a fourth value to indicate a conflict of
  interest, but we do not consider these bids here; we consider them
  separately in Section~\ref{sec:coi}.}  If a reviewer takes no action
with respect to a paper, the default bid is no.

We used the EasyChair conference management tool.  A reviewer in the
bidding process is presented with EasyChair's standard bidding page.
This shows the title of each paper on a separate row.  Single-blind
reviewers also see the authors before the paper title, with hyperlinks
to each author's homepage if available.  Institutions are not shown at
this stage.  There is a link to see details of the paper, and another
link to download the full paper itself.  The details page provides
some additional information, such as keywords, the abstract for the
paper, and for single-blind reviewers, the list of authors with
affiliations.  The PDF document itself does not list authors or their
affiliations.

The distribution of bids per reviewer is shown in
Table~\ref{tab:bid-count}.  The table shows that 60\% of reviewers
have at least 20 bids, which is a reasonable number to perform an
effective allocation of papers.  We will discuss below (in
Section~\ref{sec:bid-analysis}) the observation that single-blind
reviewers appear to enter more reviews.

\begin{table}
\begin{tabular}{l|c|c|c|c|c}
\makecell{Num\\ bids} &
\multicolumn{2}{c|}{\makecell{Single \\ blind \\ reviewers}} & 
\multicolumn{2}{c|}{\makecell{Double \\ blind \\ reviewers}} &
\makecell{Common\\1-CDF} \\ \hline
\rule{0pt}{2.6ex}
& Count & CDF & Count & CDF \\
0--4 & 6 & 5\% & 4 & 3.3\% & 100\% \\
5--9 & 9 & 12.4\% & 7 & 9.1\% & 96\% \\
10--14 & 24 & 23.2\% & 13 & 19.8\% & 89\% \\
15--19 & 21 & 49.6\% & 13 & 30.6\% & 74\% \\
20--24 & 21 & 66.9\% & 15 & 43.0\% & 60\% \\
25--29 & 25 & 87.6\% & 44 & 79.3\% & 45\% \\
30--34 & 9 & 95\% & 11 & 88.4\% & 17\% \\
$>=40$ & 6 & 100\% & 14 & 100\% & 8\% \\
\end{tabular}\vspace{1ex}
\caption{Distribution of number of bids for single-blind and double-blind
  reviewers.  1-CDF is computed over the union of single-blind and
  double-blind bids.}
\label{tab:bid-count}
\end{table}

Per the experiment design in Section~\ref{sec:exp}, we then use
EasyChair's standard tools to allocate exactly two double-blind
reviewers and two single-blind reviewers to each paper.  Once these
paper assignments are complete, each reviewer is directed to a page
listing his/her assigned papers.  This page lists the title of the
paper, with links to download the paper and see additional
information.  Single-blind reviewers see the authors here, and may
see the affiliations on the additional information page.

The submitted papers themselves are all anonymized, so author and
affiliation information is not available in the PDF document.

Due to some mid-stream withdrawls, the number of papers in
consideration at the end of the experiment was exactly 500.  Of these,
453 have four reviews and 47 have three reviews.

\subsection{Score and Rank information from reviewers}

Reviewers used a standard form to enter reviews.  This form includes
various fields for textual information, but also includes two fields
to which we pay special attention in this study: \emph{score} and
\emph{rank}.  Score represents an overall recommendation for the
paper, while rank represents a relative judgment of the paper
compared to others reviewed by the same reviewer.

\begin{table}
\begin{tabular}{l|c|p{2in}} \\
Value & Score & Description \\ \hline
\makecell*[lc]{Strong\\ accept} & 6 & \makecell*[lc]{I think this paper is well
  above \\  the bar and will fight for it} \\ \hline
\makecell*[lc]{Accept} & 3 & \makecell*[lc]{I think this paper should be \\ accepted} \\ \hline
\makecell*[lc]{Borderline} & -2 & \makecell*[lc]{I think this paper is below the
  bar, \\ but am open to accept if there \\ is strong support} \\ \hline
\makecell*[lc]{Reject} & -4 & I think this paper should be rejected \\ \hline
\makecell*[lc]{Strong\\ reject} & -6 & \makecell*[lc]{I think this paper is well
  below the \\ bar and will fight against it} \\
\end{tabular}\vspace{1ex}
\caption{Reviewers selected a score for each paper from these
  options.}
\label{tab:score-values}
\end{table}

WSDM 2017 used asymmetric scoring, so the reviewers select one of the
score values shown in Table~\ref{tab:score-values}.  Likewise,
reviewers select one of values in Table~\ref{tab:rank-values} for the
rank of the paper.  The rank values are not checked for
consistency---it is for instance possible to rank all papers as the
top paper, although we did not see such anomalies.

\begin{table}
\begin{tabular}{l|l} \\
Value & Description \\ \hline
4 & Top paper in my batch \\
3 & Top 25\% in my batch \\
2 & Top 50\% in my batch \\
1 & Bottom 50\% in my batch
\end{tabular}\vspace{1ex}
\caption{Reviewers selected a rank for each paper from these
  options.}
\label{tab:rank-values}
\end{table}

\subsection{Metadata for Implicit Bias Analysis}

For our analysis, we generate some additional metadata as part of our
exploration of the behavior of single-blind versus double-blind
reviewers.  First, we attempt to compute a country for each paper as
the plurality value of this property across the authors of the paper.
That is, if there is a single country with strictly more authors than
any other country (even if this is not a majority), we declare this to
be the country of the paper.

For each (reviewer, paper) pair, we compute the following six boolean
covariates:
\begin{enumerate}
\item Academic paper.  We hand-wrote a set of rules to determine
  whether an author's institution is academic or not (corporate,
  governmental, non-profit, unaffiliated, are all considered
  non-academic institutions).  If a strict majority of the authors are
  from an academic institution, we consider the paper to be an
  academic paper.
\item Female author.  We attempt to determine if at least one of the
  paper's authors is female.  Earlier work typically considered papers
  whose first author was female, but submissions to WSDM do not always
  follow the same conventions for first authors, so we did not have a
  reliable way to determine if one author contributed more than
  another.  Hence, we consider papers with a female author versus
  papers with no female author.\footnote{In Section~\ref{sec:female}
    we consider other alternatives to this approach.}  To make this
  determination, we manually annotated the gender of the 1491 authors.
  We found 1197 male authors, 246 female authors, and 48 authors for
  whom we could not determine gender from online searches.
\item Paper from USA.  This feature is true if the country of the
  paper as defined above is the USA.
\item Famous author.  We define a \emph{famous author} to be an author
  with at least 3 accepted papers at earlier WSDM
  conferences~\cite{wsdm}, and at least 100 papers according to dblp
  records.  There are 57 such authors.  This property is true if the
  paper has at least one famous author.
\item Same country as reviewer.  We wished to study whether knowledge
  of the authors would allow a reviewer from the same country to treat
  the paper preferentially.  This feature is true if the country of
  the paper as defined above is the same as the country of the
  reviewer as provided during the EasyChair registration process.
\item Top university.  We define top universities as the top 50
  global computer science universities.\footnote{Per
    \url{topuniversities.com}} While this choice is imperfect, the
  universities align reasonably well with our expectations for top
  universities.
\item Top companies.  We define top companies as Google, Microsoft,
  Yahoo!, and Facebook.  This property is true if any author is from a
  top company.
\end{enumerate}

Table~\ref{tab:features} gives information on the distribution for
each of these features.

\input features-tabular-bill



\subsection{Blinded Paper Quality Score} \label{sec:bpqs}

For our analysis, we need a proxy measure for the intrinsic quality of
each paper.  The rationale for this is twofold: (a) The primary task
of the reviewers is to rate paper quality and we want to represent
this null hypothesis in the feature set.  In this sense, implicit
biases would be those effects that are present above and beyond that
accounted for by the quality of the paper itself.  (b) Almost by
definition, implicit biases are second-order effects.  By directly
measuring intrinsic quality, we can reduce the background noise and
more easily detect the presence of any second-order effects.

We construct this paper quality score from the blinded raters by
combining linearly their scores and ranks, here standardized to have
zero mean and unit variance.  Among the blinded reviewers, the
correlation between these two measures is 0.75, and principal
components would combine these with equal weights.  However, we choose
to maximize the correlation between the pairs of blinded reviewers of
the same paper.  For a given score $s$ and rank $r$, this
between-reviewer correlation is maximized by a quality score $q=s +
0.111r$.  The achieved correlation between the two blinded raters is
0.38, a point to which we return in Section~\ref{sec:agreement}.

We take the quality score of a paper to be the average quality score
of the double-blind reviews for that paper, referred to below as
\texttt{bpqs}, for blinded paper quality score.  We normalize
\texttt{bqps} to have unit standard deviation.

\subsection{Bid Attractiveness Scores: Bids by Reviewer and Bids by
  Paper} \label{sec:bbr} \label{sec:bbp}

By analogy with \texttt{bpqs}, for modeling bid behavior we develop
two first-order scores.  To encode the willingness of a particular
reviewer to bid, we calculate the total bids of that reviewer; we
refer to this score as \texttt{bbr}, the bids by reviewer.  In order
to score the intrinsic bid-attractiveness of the paper, we calculate
the total number of bids on this paper by the double-blind reviewers;
we refer to this score as \texttt{bbp}, the bids by paper.  In
modeling bids, we will employ both these scores as covariates.

%% file: features-tabular-bill.tex
\begin{table}
\centerline{
\begin{tabular}{l|c|c|c}
Factor & \makecell[c]{Feature\\ name} & \makecell[c]{Number of\\ papers} &\makecell[c]{Fraction of\\ papers} \\ \hline
Paper from USA & usa & 176 & 35\% \\
\makecell[lt]{Same country\\ \hspace{1em}as reviewer} & same & 146 & 29\% \\
Female author & wom & 219 & 44\% \\
Famous author & fam & 81 & 16\% \\
Academic & aca & 370 & 74\% \\
Top university & uni & 135 & 27\% \\
Top company & com & 90 & 18\% \\
\end{tabular}}\vspace{1ex}
\caption{Summary of features and prevalence.}
\label{tab:features}
\end{table}

%% file: analysis.tex
\section{Analysis} \label{sec:analysis}

We now present our analysis of the experimental data described in
Section~\ref{sec:data}.  

\subsection{Modeling reviews} \label{sec:review-analysis}

Our modeling approach is to predict the likelihood that a single-blind
reviewer will give a positive (accept) score to a paper, using the
following multinomial logistic regression model:
$$\frac{\Pr[\mbox{score} > 0]}{\Pr[\mbox{score} <= 0]} = e^{\langle
  \Theta,v\rangle}.$$

where $\Theta$ is a set of learned parameters, and $v$ is a vector of
features consisting of a constant offset feature, the overall paper
quality score \texttt{bpqs} defined in Section~\ref{sec:bpqs} (a first-order
feature), and the six implicit bias booleans in
Table~\ref{tab:features} (second-order features).

\input review-tabular-bill


We present the results of the logistic regression in
Table~\ref{tab:review-logistic}.  There are significant non-zero
weights for the \texttt{fam} ($p=.027$), \texttt{uni} ($p=.012$) and
\texttt{com} ($p=0.002$) features.  The corresponding odds multipliers
are $1.63$ for \texttt{fam}, $1.58$ for \texttt{uni}, and $2.10$ for
\texttt{com}.  The other features do not show significant effects.

Our hypothesis in undertaking the work was that it would be very
difficult to see any effects on review behavior given the scale of the
data, and the difficulty other studies have encountered in finding
significant biases for single-blind reviewing.  Thus, we were
surprised to encounter three significant effects with substantial
odds multipliers.

The ratio of these coefficients can also be compared to the $0.80$
coefficient of \texttt{bpqs}; the result measures effect size relative
to the underlying quality of \texttt{bpqs}.  The ratio for
\texttt{fam}, \texttt{uni} and \texttt{com} correspond to shifts of
$0.61$, $0.57$, and $0.92$ standard deviations respectively.  For
\texttt{wom}, the odds multiplier of $0.78$, equivalent to $-0.31$
\texttt{bpqs} standard deviations, is not statistically significant
($p=0.16$).

\subsection{Modeling bids} \label{sec:bid-analysis}

We take a similar approach to modeling bids, but some changes are
required, as a reviewer may bid for an arbitrary number of papers.

As Table~\ref{tab:bid-count} suggests, the first question we should
reasonably ask is whether single-blind and double-blind reviewers bid
for the same number of papers.  We test this using a Mann-Whitney
test, and find that single-blind reviewers bid for more papers
(p=0.0002).  On average, single-blind reviewers bid for 19.9 papers
compared to 24.9 for double-blind reviewers, a decrease of 22\%.

Thus, the difference in behavior between the two reviewer classes is
quite significant.  We now ask a follow-on question: given that
single-blind reviewers bid more, do they bid more for particular types
of papers?  To answer this question, we pursue a similar analysis to
our regression study of review scores.  However, rather than including
an overall paper quality score (\texttt{bpqs}) into the regression, we
instead include covariates for the bid-appetite of the reviewer
(\texttt{bbr}) and the bid-attractiveness of the paper (\texttt{bbp})
as described in Section~\ref{sec:bbr}.  We retain the constant offset
term.

\input bid-tabular-bill


The results are shown in Table~\ref{tab:bid-logistic}.  In addition to
the difference in likelihood to bid, we also see that the \texttt{uni}
feature is significant ($p=0.011$), as is the \texttt{com} feature
($p=0.010$), indicating that the bids entered by single-blind
reviewers tend to favor top universities and companies, with modest
odds multipliers of $1.13$ and $1.17$ respectively.

\subsection{Additional analysis}

\subsubsection{The Matilda effect} \label{sec:female}

As described in Section~\ref{sec:rel}, there is significant work
regarding the importance of author gender in reviewing.  Some of
this work clearly points to lower assessments of scientific merit
for work purportedly authored by women.  For both bidding and
reviewing, we do not see a behavior difference between single-blind
and double-blind reviewers for papers with a female author.

We re-ran the same logistic regression analysis from two additional
perspectives: papers whose first author is female (16.4\% of papers),
and papers written by a strict majority of female authors (3.8\% of
papers).  In both cases, we do not see a significant $p$-value for the
\texttt{wom} feature.  We therefore do not see evidence that gender of
authors influences bidding or reviewing behavior.  However,
Section~\ref{sec:meta} shows meta-analysis studying our
results in the context of other results from the literature, and in
this setting we do find an overall significant gender effect.

\subsubsection{Aggregate review statistics}

We checked the lengths of reviews along with the distribution of
scores and ranks across the single-blind and double-blind conditions.
The results are shown in Table~\ref{tab:reviews}.  Average review
length for single-blind reviewers is 2073 characters versus 2061 for
double-blind, not significantly longer for either condition by
Mann-Whitney test (p=0.81).  Scores and ranks show a similar pattern,
with no significant difference in either score or rank distribution.

\begin{table}
\begin{tabular}{l|c|c|c}
Measure & \makecell[lc]{Single\\ blind \\ average} &
\makecell[lc]{Double\\blind\\average} & 
\makecell[lc]{Mann-Whitney\\$p$-value} \\ \hline
Review length & 2073 & 2061 & 0.81 \\
Reviewer score & -2.07 & -1.90 & 0.51 \\
Reviewer rank & 1.89 & 1.87 & 0.52
\end{tabular}\vspace{1ex}
\caption{Aggregate comparison of review statistics.}
\label{tab:reviews}
\end{table}

\subsubsection{Reviewer agreement} \label{sec:agreement}

A standard argument suggests that single-blind reviewers would
correlate slightly better than double-blind reviewers, for instance
because they would tend to share a preference for papers by famous
authors.

Although our study has focused on implicit biases of reviewers, the
lack of agreement among reviewers is also notable.  In part, this can
be mitigated by using more than one reviewer.  The inter-rater
reliability associated with an average of $n$ raters sharing
correlation $\rho$ has correlation $\sqrt{nR/(1+nR)}$ where
$R=\rho^2/(1-\rho^2)$ is the ratio of explained to unexplained
variance.

For example, the inter-reviewer correlation for \texttt{bpqs} is 0.38,
which corresponds to having benefit and inaccuracy of $n=1$ reviewer
per paper.  Under our current protocol we have $n=2$ blinded
reviewers, and the operative correlation is 0.5.  Were WSDM to enact
double-blinded review, we would have at our disposal $n=4$ reviewers
and the operative correlation would be 0.63, while $n=5$ and 6 achieve
correlations of 0.68 and 0.71 respectively.  Correlations of 0.6
characterize imperfect human-based measurement systems, and are common
enough in contexts where the low-value material has been excluded from
human assessment.

In summary, we recommend that conference organizers be cognizant of
the inter-reviewer agreement that their review process provides, and
choose appropriately the number of reviews that each paper receives.





\subsubsection{Changes during discussion} \label{sec:after}

Finally, we may ask what happens after the experiment concludes and
the discussion phase begins.  During this phase is it common to see
some changes in review scores.  We analyzed these scores, and saw 32
changes to scores entered by single-blind reviewers compared to 41
changes to scores entered by double-blind reviewers.  This difference
is not significant (Fisher-Exact, p=0.28).  We compared the changes in
scores to determine whether double-blind reviewers tend to have
changes of larger magnitude than single-blind reviewers.  The
distributions of score changes are not significantly different
(Mann-Whitney, p=0.58).  We then checked whether double-blind
reviewers tend to move more in the direction of the initial mean score
than single-blind reviewers after discovering the authors of the
paper.  Here also, we find no difference in the magnitude of shifts
towards the mean (Mann-Whitney, p=0.58).  Hence, during the discussion
phase, after the authors have been revealed, we cannot conclude that
the initially double-blind reviewers behave differently from
single-blind reviewers.

\subsubsection{Conflicts of interest} \label{sec:coi}

It is natural to hypothesize that in a double-blind setting there will
be fewer declared conflicts of interest, as reviewers will not
recognize possible conflicts.  In WSDM 2017, the EasyChair tool
automatically (but imperfectly) detects conflicts based on the email
domains of authors and reviewers.  Reviewers may specify additional
conflicts as they bid for papers.  It is possible to configure the
system to allow authors to specify conflicts with PC members at
submission time, but we did not enable this configuration.

We consider the overall set of conflicts generated both automatically
by EasyChair and by reviewer specification.  We find that the total
number of reviewers expressing a conflict (59/121 in the single-blind
setting versus 47/121 in the double-blind setting) is not
significantly different (Fisher-Exact, p=0.35).  Likewise, the number
of conflicts expressed by those reviewers who express a conflict is
not significant (Mann-Whitney, p=0.63).  Hence, in the settings we
adopted, we do not see that double-blind reviewing introduces a
significant difference in expression of conflicts of interest.

\subsection{Discussion}

There are several questions one may raise with respect to our
experiment.  First is the issue that we study the behavior of the PC
with respect to bidding and scoring papers only.  After these steps
are complete, the SPC member conducts some discussion among the
reviewers, and the program chairs make a final decision.  While
Section~\ref{sec:after} suggests there may not be significant changes
specifically in how reviewers modify their scores during discussion,
it is nonetheless possible that during these stages, the final
acceptance decision may show unexpected behaviors.  This is clearly an
area for further work.  However, we have observed that the critical
inputs to this final decision stage (score and rank of reviewers) are
impacted significantly by the reviewing model.

It is possible also that PC members behaved differently in our setting
than they would in a ``pure'' reviewing situation involving only a
single type of reviewing.  For instance, single-blind reviewers in our
experiment were nonetheless presented with papers that do not include
author names and affiliations.  We also mentioned briefly in the
call for papers that we would experiment with double-blind reviewing
this year.  We do not have a rigorous methodology to estimate the
nature of these biases, but we observe that at least in the case of
presenting anonymized PDFs to both types of reviewers, it is plausible
that this would cause us to under-estimate rather than over-estimate
the effects.

Having stated those caveats, we now discuss some issues with respect
to the practical implementation of double-blind reviewing.

\subsubsection{Practical issues with double-blind reviewing} \label{sec:impl}

There is a long-standing question whether it is practical to anonymize
a submission.  This question depends on the nature of the field (for
instance, it would be impossible to anonymize work in a large and
well-known systems project).  Hill \emph{et al.}~\cite{hill03} argue
that it is possible to automatically identify authors in many cases
based on the text of the paper alone.  However, other studies have
observed that reviewers' guesses about authorship are often
wrong~\cite{snodgrass2006}.

A second issue in the practical difficulty of retaining anonymity in
double-blind reviewing is the increasingly common practice of
publishing early versions of work on \texttt{arXiv.org}.  For example,
this paper appeared on arXiv before being submitted to any
peer-reviewed venue.  This practice was a significant contributor to
the decision of the Journal of the American Economic Association to
abandon double-blind reviewing~\cite{jaschik11}.  WSDM 2017 did not
state a policy with regard to publishing pre-prints on arXiv, but when
asked, we discouraged but did not forbid such publication.  In its
2016 call for papers~\cite{nips2016}, the NIPS machine learning
conference, which performs double-blind reviewing, informed authors
that prior submissions on arXiv are allowed, but reviewers are asked
``not to actively look for such submissions.''  If reviewers happened
to be aware of the work, NIPS nonetheless allows the reviewing to
proceed.

These practical issues appear to be significant and unresolved.

%% file: review-tabular-bill.tex
\begin{table}
\begin{tabular}{l|r|r|c|r|r|c}
Name & Coeff. & Stderr & \makecell[c]{Conf.\\interval} & p-value & \makecell[c]{Odds\\mult.} &  \makecell[c]{\texttt{bpqs}\\equiv.} \\ \hline
const & -1.83 & 0.24 & [-2.31,-1.36] & 0.000 & 0.16 & - \\
bpqs & 0.80 & 0.08 & [0.64,0.97] & 0.000 & 2.23 & 1.00 \\
com & 0.74 & 0.24 & [0.27,1.21] & 0.002 & 2.10 & 0.92 \\
fam & 0.49 & 0.22 & [0.05,0.93] & 0.027 & 1.63 & 0.61 \\
uni & 0.46 & 0.18 & [0.09,0.83] & 0.012 & 1.58 & 0.57 \\
wom & -0.25 & 0.18 & [-0.60,0.10] & 0.160 & 0.78 & -0.31 \\
same & 0.14 & 0.24 & [-0.34,0.62] & 0.564 & 1.15 & 0.17 \\
aca & 0.06 & 0.22 & [-0.38,0.51] & 0.775 & 1.07 & 0.08 \\
usa & 0.01 & 0.21 & [-0.42,0.44] & 0.964 & 1.01 & 0.01 \\
\end{tabular}
\vspace{1ex}
\caption{Learned coefficients and significance for review score
  prediction.}
\label{tab:review-logistic}
\end{table}

%% file: bid-tabular-bill.tex
\begin{table}
\begin{tabular}{l|r|r|c|r|r}
Name & Coeff. & Stderr & \makecell[c]{Conf.\\interval} & p-value & \makecell[c]{Odds\\mult.} \\ \hline
const & -4.87 & 0.08 & [-5.04,-4.71] & 0.000 & 0.01 \\
bbr & 0.05 & 0.00 & [0.04,0.05] & 0.000 & 1.05 \\
bbp & 0.08 & 0.00 & [0.07,0.09] & 0.000 & 1.09 \\
com & 0.16 & 0.06 & [0.04,0.28] & 0.010 & 1.17 \\
uni & 0.12 & 0.05 & [0.03,0.22] & 0.011 & 1.13 \\
fam & 0.07 & 0.06 & [-0.06,0.19] & 0.287 & 1.07 \\
wom & 0.05 & 0.04 & [-0.04,0.14] & 0.268 & 1.05 \\
usa & 0.02 & 0.05 & [-0.07,0.11] & 0.681 & 1.02 \\
aca & 0.01 & 0.06 & [-0.10,0.12] & 0.881 & 1.01 \\
\end{tabular}
\vspace{1ex}
\caption{Learned coefficients and significance for bid prediction.}
\label{tab:bid-logistic}
\end{table}

%% file: meta.tex
\section{Meta-Analysis} \label{sec:meta}

\providecommand{\okike}{Okike \emph{et al.}}
\providecommand{\knoblock}{Knoblock-Westerwick \emph{et al.}}
\providecommand{\tung}{Tung}
\providecommand{\budden}{Budden \emph{et al.}}
\providecommand{\roberts}{Roberts and Verhoef}
\providecommand{\blank}{Blank}
\providecommand{\tomkins}{Tomkins \emph{et al.}}

In this section, we compare the effect sizes reported in
Table~\ref{tab:review-logistic} to those reviewed in
Section~\ref{sec:rel}.  We focus on 6 empirical studies:
\tung~\cite{tung06}, \knoblock~\cite{matilda13},
\budden~\cite{budden2008}, \blank~\cite{blank91}, \okike~\cite{jama16}
and \roberts~\cite{evolang16}.  By including our work above, we have 7
studies total.

We report all effect sizes as log-odds multipliers. This choice allows
direct use of Table~\ref{tab:review-logistic}'s logistic regression
coefficients, involves modest recalculations for 3 of the other 6
studies, and is reasonably interpretable. Two studies, \knoblock\ and
\roberts, report $t$-statistics. For \tung, we have the annual
aggregates, so we can recover the $t$-statistic relative to the
binomial distribution. The method of Hasselblad and
Hedges~\cite{hasselblad95} allows us to transform $t$-statistics into
location shifts of a continuous logistic distribution, The result is
interpretable on the log-odds scale.

\begin{table}

\begin{tabular}{lllrr} \hline
Source & \makecell[lc]{Study \\ context} & 
Method & \makecell[lc]{Effect \\ size} & \makecell[lc]{Effect\\rank}
\\ \hline
\hline
\multicolumn{5}{c}{Famous author} \\ \hline
Tung & SIGMOD & $t\rightarrow$ log odds & 1.619 & 1 \\ \hline
& VLDB & $t\rightarrow$ log odds & 1.306 & 2 \\ \hline
\okike & CORR & log OR & 1.136 & 3 \\ \hline
\tomkins & WSDM & logistic reg & 0.506 & 4 \\ \hline
\hline
\multicolumn{5}{c}{Female author} \\ \hline
\budden & BE & log OR & -0.426 & 2 \\ \hline
& BES & log OR & -0.119 & 10 \\ \hline
& AB & log OR & -0.224 & 8 \\ \hline
& BC & log OR & -0.398 & 4 \\ \hline
& JB & log OR & -0.048 & 11 \\ \hline
& LE & log OR & -0.370 & 5 \\ \hline
& \makecell[lc]{all 6 \\ journals} & log OR & -0.246 & 6 \\ \hline
\makecell[lc]{Knoblock-\\Westerwick\\\emph{et al.}} & 2010 ICA & $t\rightarrow$ log odds & -0.410 & 3 \\ \hline
\blank & AER & log OR & -0.229 & 9 \\ \hline
\makecell[lc]{Roberts\\and\\Verhoef} & EvoLang 11 &  $t\rightarrow$ log odds & -1.186 & 1 \\ \hline
\tomkins & WSDM & logistic reg & -0.25 & 7 \\ \hline
\hline
\multicolumn{5}{c}{Top 50 institutes} \\ \hline
\blank & AER & log OR & 0.020 & 2 \\ \hline
\tomkins & WSDM & logistic reg & 0.46 & 1 \\ \hline
\end{tabular}
\caption{Across seven studies and 14 contexts, 17 effects as log-odds
  multipliers.}
\label{tab:meta}
\end{table}

Table~\ref{tab:meta} presents all effects and their ranks. For famous
authors and top 50 institutions, the number of studies is small enough
that no study can be called an outlier. That said, for famous authors,
our value of 0.51 is actually smaller than that reported elsewhere,
while for top 50 institutions, our value of $0.46$ is larger.

For the effect of female authors, the pattern is almost uniformly
negative. The -$0.25$ effect reported in
Table~\ref{tab:review-logistic} ranks 7 out of 11, on the small side,
albeit this value is not qualitatively different from the -$0.246$
combined effect of \budden\ and the $-0.229$ effect of Blank. By the
standards of meta-analysis, in aggregate, the effect against females
authors can be considered statistically significant, albeit with
continuing caveats regarding the observational nature of some of these
studies.

In summary, the famous author effect we report is reported by others
as even larger. For the effect of female authors, we report a value
that is in line but somewhat smaller. Two interpretations suggest
themselves: First, we may be observing the natural variation among
such studies. Alternately, relative to journal reviews, the conference
review process may operate with slightly different biases, and social
affiliation (famous authors, top 50 institutions, and top companies)
may play an enhanced role. Future research should clarify this.

%% file: conclusion.tex
\section{Conclusion} \label{sec:conclusion}

In conclusion, the heart of our findings is that single-blind
reviewers make use of information about authors and institutions.
Specifically, single-blind reviewers bid less yet are differentially
more likely to bid on papers from top institutions, and more likely to
recommend for acceptance papers from famous authors or top
institutions, compared to their double-blind counterparts.  Regarding
the gender effect the situation is more nuanced.  Our results do not
show a statistically significant effect for gender, but our
meta-analysis places our findings in line with other experiments,
which in aggregate warrant a conclusion that the gender effect is
significant.

The primary ethical question is whether this behavior is okay.  In one
interpretation, single-blind reviewers make use of prior information
that may allow them to make better overall judgments.  As a
consequence, however, it may be that other work is disadvantaged, in
the sense that two contributions of roughly equal merit might be
scored differently by single-blind reviewers, in favor of the one from
a top school, while double-blind reviewers may not show this bias as
strongly.

Clearly, our understanding of the implications of reviewing
methodologies remains nascent.  We feel that program and general
chairs of conferences should seriously consider the advantages of
employing double-blind reviewing.  Furthermore, we recommend that
conferences quantify and remain cognizant of inter-reviewer agreement.

%% file: appendix.tex
\appendix

\section{Appendix}

\subsection{Mechanism of Running the Experiment}

Our experiment was performed using capabilities that already exist in
EasyChair, plus a few last-minute changes provided by the EasyChair
team.  The workflow within the EasyChair system to perform the
experiment relies on the use of a little-used reviewing model known as
``External Review Committee,'' and runs as follows:
\begin{enumerate}
\item Disable subreviewers.
\item Change reviewing model to ERC.  
\item Configure PC to see author names and ERC as double-blind.
\item Configure access to reviews for both PC and ERC as ``see only their
own reviews.''
\item Invite randomly split half of reviewers
  into PC and other half into ERC.
\item Invite or add senior as standard PC members
\item Receive papers, perform do paper bidding.
\item For Senior PC members and program chairs, configure at most 0 papers assigned.
\item Configure 2 papers per member for regular PC and ERC members.
\item Run automatic paper assignment separately for PC and ERC.
\item Assign papers to senior using interactive paper assignment.
\item Run standard reviewing period for both PC and ERC.
\item When reviewing is over, change all ERC members to standard PC members
\item Change reviewing model from ERC to Senior PC.
\item Change all SPC members from regular PC to SPC in the new
  reviewing model.
\item Change review access to PC members can see all reviews for their
  papers.
\item From now on, run discussion and decision process using standard flows.
\end{enumerate}